\documentclass[twocolumn]{aa}
\usepackage{graphicx}
\usepackage{astrobib}
\usepackage{journals}

\usepackage{txfonts}
\begin{document}
\newcommand\degd{\ifmmode^{\circ}\!\!\!.\,\else$^{\circ}\!\!\!.\,$\fi}
\newcommand{\rdm}{{\rm\ rad\ m^{-2}}}

   \title{Radio Linear and Circular Polarization from M81*}

   \author{Andreas Brunthaler\inst{1,2},
           Geoffrey C. Bower\inst{3}, 
           Heino Falcke\inst{4,5}
           }

   \institute{Joint Institute for VLBI in Europe, Postbus 2, 7990 AA
              Dwingeloo, The Netherlands
              \and
              Max-Planck-Institut f\"ur Radioastronomie, Auf dem H\"ugel 69,
              53121 Bonn, Germany
              \and
	      Radio Astronomy Laboratory, University of California,
	      Berkeley, CA 94720, USA
              \and    
	      ASTRON, Postbus 2, 7990 AA Dwingeloo, The Netherlands
	      \and
	      Department of Astrophysics, Radboud Universiteit
              Nijmegen, Postbus 9010, 6500 GL Nijmegen, The Netherlands}

   \offprints{brunthal@mpifr-bonn.mpg.de}
   \date{Received 30 December 2005 / Accepted 19 January 2006}

   \abstract{We present results from archival Very Large Array (VLA) data 
             and new VLA observations to investigate the long term behavior of 
             the circular polarization of M81*, the nuclear radio source
             in the nearby galaxy M81. We also used the 
             Berkeley-Illinois-Maryland Association (BIMA) array to observe 
             M81* at 86 and 230 GHz.  M81* is unpolarized in the 
             linear sense at a frequency as high as 86 GHz
             and  shows variable circular polarization at a frequency as 
             high as 15 GHz.  The spectrum of the fractional circular
	     polarization is inverted in most of our observations.  The
	     sign of circular polarization is constant over frequency and time.
             The absence of linear polarization sets a lower limit to
             the accretion rate of $10^{-7} M_\odot y^{-1}$.
             The polarization properties are strikingly similar to the 
             properties of Sgr A*, the central radio source in the Milky Way. 
	     This supports the hypothesis that M81* is a scaled up version of 
             Sgr A*. On the other hand, the broad band total intensity spectrum
             declines towards milimeter wavelengths which differs from previous
	     observations of M81* and also from Sgr A*.
 
   }

   \maketitle
%
%________________________________________________________________

\section{Introduction}

The nearby spiral galaxy M81 (NGC\,3031) shares many properties with
the Milky Way. It is similar in type, size and mass and it also
contains a nuclear radio source, M81*, that is most likely associated
with a supermassive black hole. M81* has been studied extensively using
Very Long Baseline Interferometry (VLBI) in the
past. \citeN{BietenholzBartelRupen2000} resolved M81* into a stationary
core with a one sided jet. Multi-wavelength
(\citeNP{HoFilippenkoSargent1996}) and sub-millimeter observations
  (\citeNP{ReuterLesch1996}) showed many similarities between M81* and
  Sgr A*, the central radio source in our Milky Way (\citeNP{MeliaFalcke2001}).
A jet model of Sgr A* has been applied to M81*, where it can reproduce the 
radio flux density and the size of the radio core by changing the accretion 
rate (\citeNP{Falcke1996}). The sizes of both radio sources show a $\sim 1/\nu$
dependency on the frequency (e.g. \citeNP{BietenholzBartelRupen2004}
for M81*, and \citeNP{BowerFalckeHerrnstein2004} and
\citeNP{ShenLoLiang2005} for Sgr A*). 

M81* is an apparent transitional object between Sgr A* and high
luminosity AGN.  As the brightest of the nearby low
luminosity AGN (LLAGN), it is 5 orders of magnitude brighter than Sgr A* at
radio wavelengths.  M81* is substantially underluminous at X-ray
wavelengths ($L\sim 10^{-5} L_{edd}$), yet not as much as Sgr A*
($L\sim
10^{-10} L_{edd}$). Still, it is the faintest LLAGN we can study.

Furthermore, the polarization properties of M81* and Sgr A* are very similar. 
Sgr A* shows circular polarization in absence of linear polarization
(\citeNP{BowerFalckeBacker1999}, \citeNP{BowerBackerZhao1999},
\citeyearNP{BowerWrightBacker1999}) and we detected the same behaviour in M81* 
(\citeNP{BrunthalerBowerFalcke2001}.)
The polarization properties of Sgr A* and M81* are in contrast to the
properties of most radio jets in active galactic nuclei where linear
polarization often exceeds circular polarization by a large factor
(e.g.~\citeNP{WardleHomanOjha1998}; \citeNP{RaynerNorrisSault2000}).
The absence of linear polarized emission in Sgr A* can be explained as a
consequence of the accretion flow.

\citeN{BowerFalckeSault2002} investigated the long term behavior of the
circular polarization in Sgr A* from archival VLA
data and showed that {\it i)} the circular polarization is variable on 
timescales of days to months, {\it ii)} the sign of the circular polarization 
stayed constant over the entire time range of almost 20 years, and {\it iii)} 
the average spectrum of circular polarization is inverted. 

After the discovery of circular polarization in M81* we used the VLA to 
investigate the variability of the circular polarization on short timescales.
We used additional archival VLA data to investigate the long term
behavior of the circular polarization of M81*.

\section{Observations}

\begin{figure*}
\resizebox{\hsize}{!}{\includegraphics[bbllx=5.5cm,bburx=19.0cm,bblly=2.2cm,bbury=27.0cm,clip=,angle=-90]{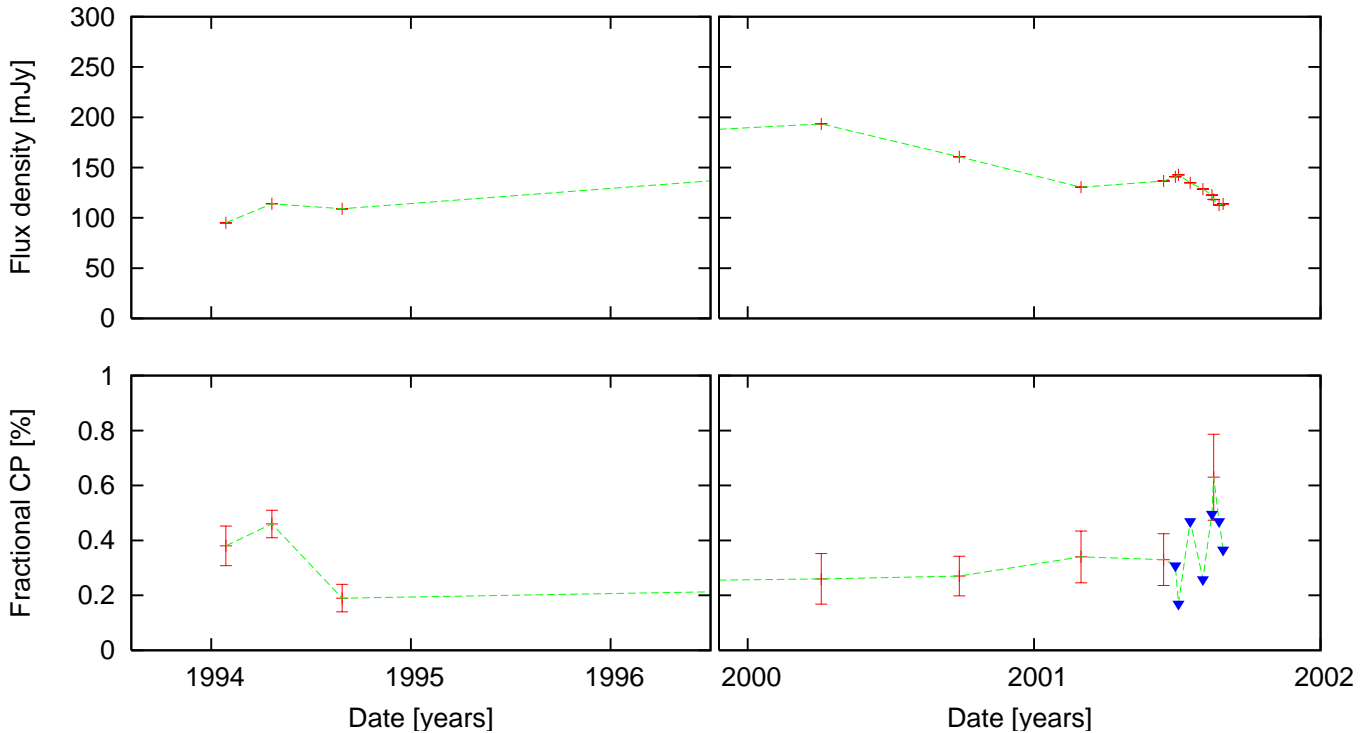}}
\caption{Light-curve in total intensity (top) and fractional circular
  polarization of M81* at 4.8 GHz for the archival data (left) and the new
  observations (right). The filled triangles are 3$\sigma$ upper limits. The 
  two data points in 2000 are taken from 
  \protect\citeN{BrunthalerBowerFalcke2001}}
\label{c-light}
\end{figure*}

\subsection{Archival VLA Data}

M81* and the supernova SN1993J in M81 were observed many
times with the Very Long Baseline Array (VLBA) and the phased Very
Large Array (VLA) over the last decade
(e.g. \citeNP{BartelBietenholzRupen1994};
\citeNP{BietenholzBartelRupen2000}). The observations were typically 16
hours in duration, were made at different frequencies and involved rapid
switching between M81*, SN1993J and scans roughly every hour on the
extragalactic background source 0954+658 for calibration purposes. We used 
the VLA data from the observations on 5 November 1993 ($\nu=$ 8.4 and 15 GHz), 
16 December 1993 (8.4 and 15 GHz), 29 January 1994 (4.8 and 8.4 GHz), 
21 April 1994 (4.8 GHz), 29 August 1994 (4.8 and 8.4 GHz), 31 October 1994 
(8.4 GHz), 23 December 1994 (8.4 GHz), and 7 April 1996 (8.4 GHz). The VLA 
data had 50 MHz of bandwidth in two sidebands in right (RCP) and left (LCP)
circular polarization modes.

Data reduction was performed with the Astronomical Image Processing
System (AIPS). 3C\,48 was used as primary amplitude calibrator. Then
amplitude and phase self-calibration was performed on 0954+658. This
forces 0954+658 to have zero circular polarization. The amplitude
calibration was transfered to M81* and SN1993J before we performed phase
self-calibration on M81* and SN1993J. Finally, we mapped all three sources
in Stokes I and V. Flux densities were determined by fitting an
elliptical Gaussian component to the sources. 

\subsection{New VLA observations}

In addition to the archival VLA data, we used the VLA to observe M81*  on 02 March 2001 and 9
dates between 15 June and 30 August 2001.  During the latter period,
the minimum and maximum separations between observing dates were 2 and
15 days, respectively.   The VLA was in B configuration for the first
observation and C array for the remaining observations.  Observations
were made at 5.0, 8.4, 15 and 22 GHz with 50 MHz of bandwidth in  two
sidebands in RCP and LCP modes.   

Each observation was between two and four hours in duration.  Observing
and analysis was performed with AIPS and followed the procedures outlined in
\citeN{BrunthalerBowerFalcke2001}. The extragalactic point source
J1044+719 was used as a phase, amplitude and polarization calibrator as
well as reference pointing source. The extragalactic point source
J1053+704 was used to check for polarization calibration errors .  3C 286 was
used as an absolute amplitude calibration source.  Finally, we mapped
all three sources in Stokes I, U, Q, and V. Flux densities were
determined by fitting an elliptical Gaussian component to the sources. 

We also used the VLA to observe M81* on 09 August 2003 at 15 GHz, 22 GHz, and 
43 GHz in polarimetric mode.  The VLA was in A configuration with a resolution 
of about 50 milliarcseconds at 43 GHz. The total integration time on M81* was 
8, 48, and 52 minutes at 15, 22 and 43 GHz respectively and spread over a 
time range of 11 hours. Phase and amplitude calibration were 
performed using the nearby compact source, J1056+714. Phase self-calibration 
was also performed on M81* to eliminate the effects of atmospheric 
decorrelation. Polarization leakage calibration was performed using 
simultaneous full track observations of J1056+714 and J1048+701.  

\begin{figure*}
\resizebox{\hsize}{!}{\includegraphics[bbllx=5.5cm,bburx=19.0cm,bblly=2.2cm,bbury=27.0cm,clip=,angle=-90]{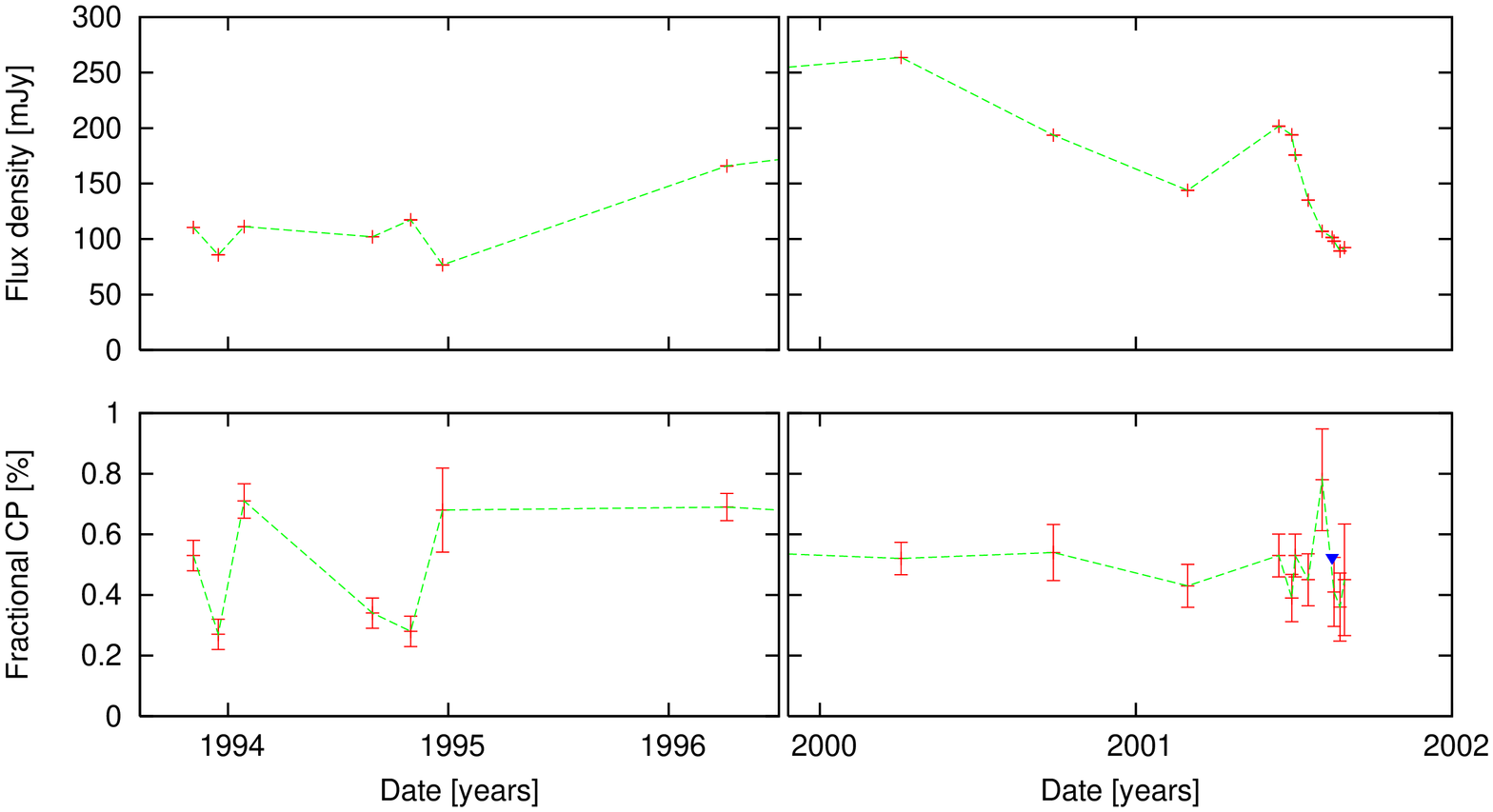}}
\caption{Light-curve in total intensity (top) and fractional circular
  polarization of M81* at 8.4 GHz for the archival data (left) and the new
  observations (right). The filled triangle is an 3$\sigma$ upper limit. The two data points in 2000 are taken from \protect\citeN{BrunthalerBowerFalcke2001}}
\label{x-light}
\end{figure*}

\begin{figure*}
\resizebox{\hsize}{!}{\includegraphics[bbllx=5.5cm,bburx=19.0cm,bblly=2.2cm,bbury=27.0cm,clip=,angle=-90]{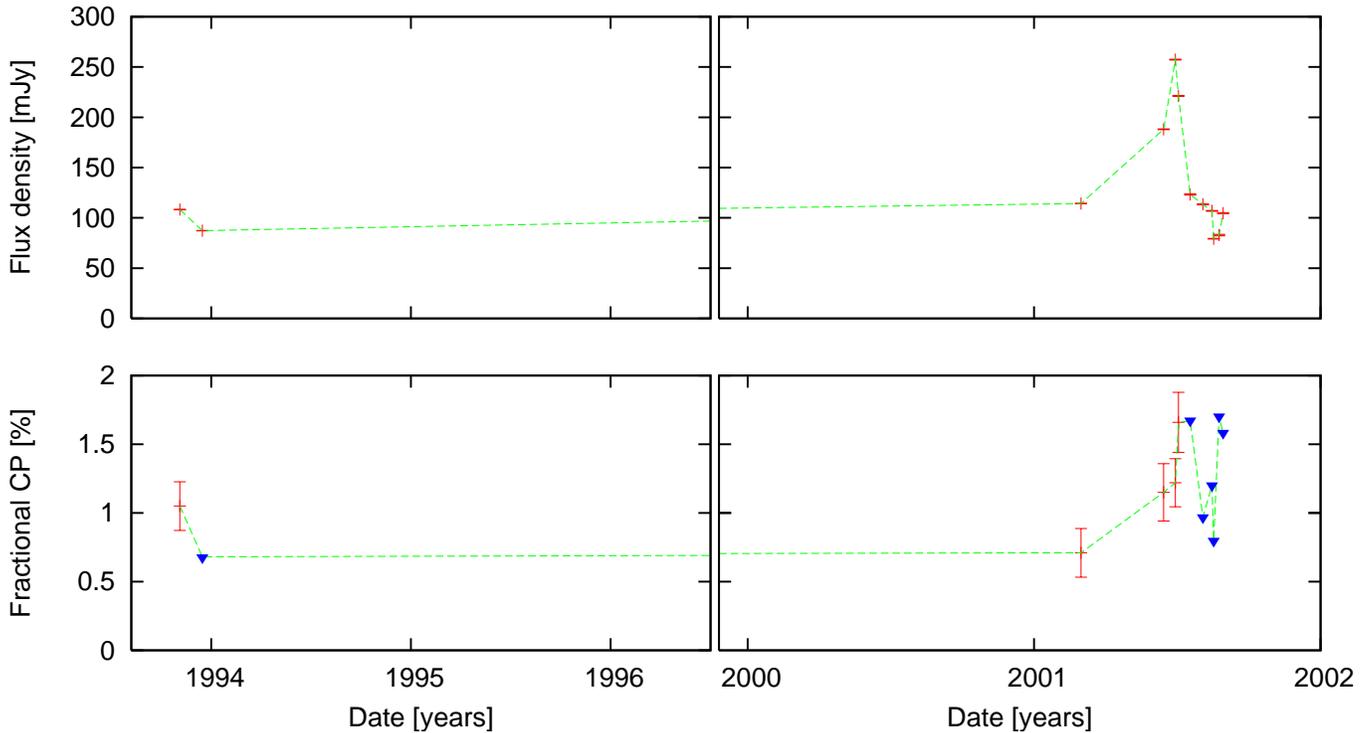}}
\caption{Light-curve in total intensity (top) and fractional circular
  polarization of M81* at 15 GHz for the archival data (left) and the new
  observations (right). The filled triangles are 3$\sigma$ upper limits.}
\label{u-light}
\end{figure*}

\subsection{BIMA observations}

We used the Berkeley-Illinois-Maryland Association (BIMA) array to
observe M81* at 3\,mm wavelength (\citeNP{WelchThorntonPlambeck1996}).  
Observations were made in the multiplexed
polarimetric mode described in \citeN{BowerWrightBacker1999}.  The receivers
were tuned to sky frequencies of 82.8 GHz (lower sideband) and
86.2 GHz (upper sideband).   The compact source 0954+658 was 
used for phase calibration.  Leakage calibration was determined
from observations of 3C 279 on 11 November 2003.  Observations of M81* were
made on 6 dates in September and October 2003 (Table~\ref{High}).
BIMA observations at 230 GHz on M81* were also performed in November 2003 (Table~\ref{High}).
These data were also obtained in polarimetric mode with similar observing
parameters to the 3\,mm observations. Each observation was a 8 hour track.

\subsection{Error Analysis}

The Stokes parameter V is measured as the difference between the left- and
right-handed parallel polarization correlated visibilities. Errors
in circular polarization measurements with the VLA have numerous origins:
thermal noise, gain errors, beam squint, second-order leakage corrections,
unknown calibrator polarization, background noise and radio frequency
interference. The errors caused by amplitude calibration errors, beam
squint, and polarization leakage scale with the source strength, and
therefore the fractional circular polarization is a more relevant
indicator for the detection of circular polarization.
A detailed discussion of these errors is given in
\citeN{BowerFalckeBacker1999}  and \citeN{BowerFalckeSault2002}. 
We calculated the systematic errors based on the model for the VLA for
circular polarization from \citeN{BowerFalckeSault2002}. For M81*,
SN1993J, and J1053+704 the errors on the fractional circular
polarization in Tables~\ref{all-c} -- \ref{all-u} are separated into
statistical and systematic terms, while for the calibrator sources
0954+658 and J1044+719 only the statistical error is given. The calibrator 
sources do not have a systematic error, since their circular polarization was
assumed to be zero during the calibration.

\section{Results}

\subsection{Circular and Linear Polarization}

The results for the archival data at 4.8, 8.4, and 15 GHz are shown in
Tables~\ref{all-c}, \ref{all-x}, and \ref{all-u} respectively. We consider 
circular polarization as detected if the measured flux density exceeds the 
combined statistical and systematic errors (added in quadrature) by a factor 
of three. 0954+658 showed no circular polarization as
expected. M81* showed circular polarization in all observations except
one (16 December 1993 at 15 GHz). SN1993J showed
circular polarization only in one epoch (29 August 1994 at 4.8 GHz)
and no circular polarization in all other observations. The upper
limits on the circular polarization of SN1993J are not very meaningful
at 15 GHz due to the low flux density and unexpected high noise. 

The results for the new observations at 4.8, 8.4, and 15 GHz are shown
in Tables~\ref{all-c}, \ref{all-x}, and \ref{all-u} respectively. The
22 GHz data and the high frequency observations on 9 August 2003 gave no useful 
limits on the circular polarization, mainly
because the short integration time and higher systematic errors. At
4.8. 8.4, and 15 GHz, 1044+719 showed no circular polarization as
expected. M81* showed circular polarization in three epochs at 4.8 GHz,
all except one epoch at 8.4 GHz and four epochs at 15 GHz. The check
source 1053+704 showed only circular polarization in two epochs at 15
GHz.

The three {\it detections} of circular polarization in the check sources
SN1993J and 1053+704 are caused by either remaining amplitude
calibration errors or by a small level of circular polarization in the
sources. Since both sources show no circular polarization at 8.4 GHz,
the frequency band with the highest sensitivity, it is most likely that
the {\it detected} circular polarization in SN1993J and 1053+704 comes
from residual amplitude calibration errors. The fractional circular
polarization in M81* at 15 GHz is higher by a factor of 2 and 4 than in
1053+704 in the observations on 15 June 2001 and 4 July 2001
respectively. In these two cases, the measured circular polarization of M81* 
is probably only partly caused by the amplitude calibration errors. 

In the BIMA observations at 86 and 230 GHz neither linear nor circular 
polarization is detected for M81* in any individual epoch. Mean linear 
polarization at 86 GHz is $1.2$ mJy, or 1.6\% ($3\sigma$). Mean circular 
polarization at 86 GHz is $2.8 \pm 0.4$ mJy, or $3.9 \pm 0.5$\%.  
Although this is formally a detection, it is not clear 
whether systematic errors are significant. Due to the low flux density, limits 
on the polarized flux density are not significant at 230 GHz. 

The 8.4 GHz data seems to be the most reliable data since M81* showed
circular polarization in all epochs except one while the check sources 
SN1993J and 1053+704 never showed circular polarization. The light-curve of 
total intensity and fractional circular polarization is shown in 
Fig.~\ref{x-light}. The fractional circular polarization shows significant 
variability on timescales of a few weeks which is not correlated with the 
variability in the total intensity. Between 4 August 2001 and 16 August 2001, 
the fractional circular polarization at 8.4 GHz dropped from 0.78\% to less 
than 0.35\% while the total intensity showed no significant change.
However it is remarkable that, despite 
the strong variability, the sign of the circular polarization is always
positive. At the other two frequencies, the sign is also positive
when circular polarization is detected in M81*.

The spectrum of the fractional circular polarization is  inverted 
($\alpha > 0$ for $m_{c} \propto \nu^\alpha$) with values between 0.4 and 2 
in most epochs when it was detected at more than one frequency. The mean 
spectral index between 4.8 and 8.4 GHz is 0.51, while the mean spectral 
index between 8.4 and 15 GHz is 1.52. Only the 
observation on 18 August 2001 shows a steep spectrum of $\alpha$=-0.77 
between 4.8 and 8.4 GHz..

\begin{figure}
\resizebox{\hsize}{!}{\includegraphics[angle=-90]{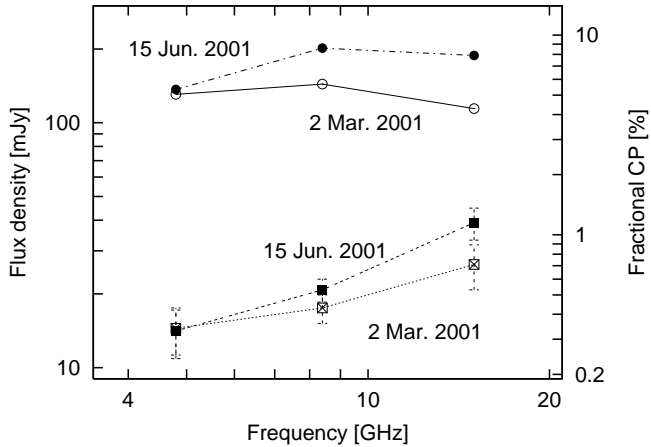}}
\caption{Spectra of the total intensity (circles) and the fractional circular polarization (squares) from 2 March 2001 (open) and 15 June 2001 (filled).
}
\label{spec}
\end{figure}

\begin{figure}
\resizebox{\hsize}{!}{\includegraphics[angle=-90]{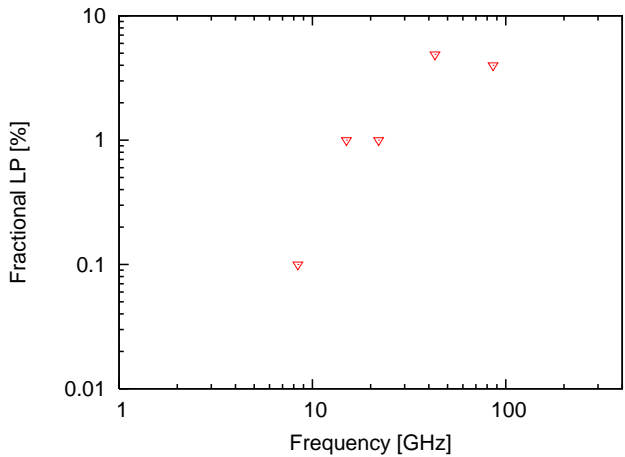}}
\caption{3 $\sigma$ upper limits on the linear polarization of M81* up to 86 GHz. The 8.4 GHz data point is taken from \protect\citeN{BowerFalckeMellon2002}.}
\label{lp}
\end{figure}

Linear polarization was not detected in the VLA observations on 9 August 2003
and the BIMA observations at 86 GHz (Fig.~\ref{lp}). 
The archival data was not searched for linear polarization.

\subsection{Total intensity}

The high frequency VLA data taken on 9 August 2003 give a simultaneous total 
intensity spectrum of M81* that shows the flux density decreasing from 15 to 
43 GHz with a spectral index of $\sim -0.6$. In the BIMA observations M81* is 
clearly detected at 86 GHz in total intensity and is strongly variable on a 
time scale $\sim 10$ days. The mean total intensity is 71 mJy. Due to low 
sensitivity, M81* is only marginally detected at 230 GHz in total intensity. 
The mean flux density for M81* from this observation is $31 \pm 4 \pm 15$ mJy, 
where the first error is the statistical error and the second error is the 
systematic error due to decorrelation. Atmospheric decorrelation may be 
serious and these results may significantly underestimate the total flux 
density of M81* (Table~\ref{High}).

\begin{figure}
\resizebox{\hsize}{!}{\includegraphics[angle=-90]{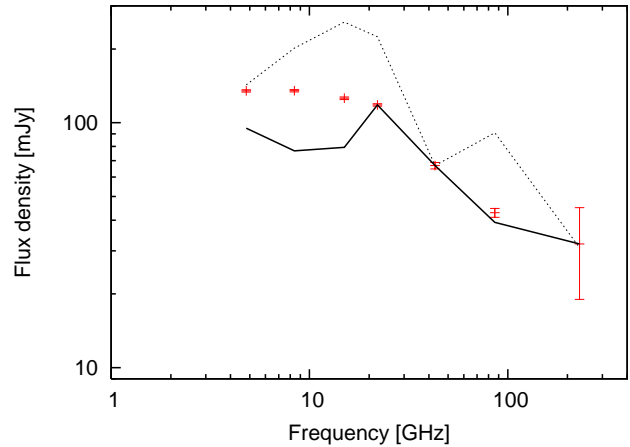}}
\caption{Non simultaneous broad-band spectrum of M81*.  Also shown are the maximal (dashed line) and minimal (solid line) measured values at each frequency.}
\label{bb-spec}
\end{figure}

Fig.~\ref{bb-spec} shows a non simultaneous broad-band spectrum of M81*. The 
15 -- 43 GHz data points are from the VLA observations on 9 August 2003. The 
4.8 and 8.4 GHz data points are from the observation on 19 July 2001, where 
the 15 GHz flux was comparable to the 15 GHz flux in the 9 August 2003 
observation. The 86 GHz and 230 GHz data points are from the BIMA observations 
on 7 September 2003 and November 2003 respectively. Also shown are the maximal 
and minimal measured values at each frequency in our observations. Although 
the spectrum is not simultaneous it is clear that the spectrum declines 
towards higher frequencies.

\begin{figure}
\resizebox{\hsize}{!}{\includegraphics[angle=-90]{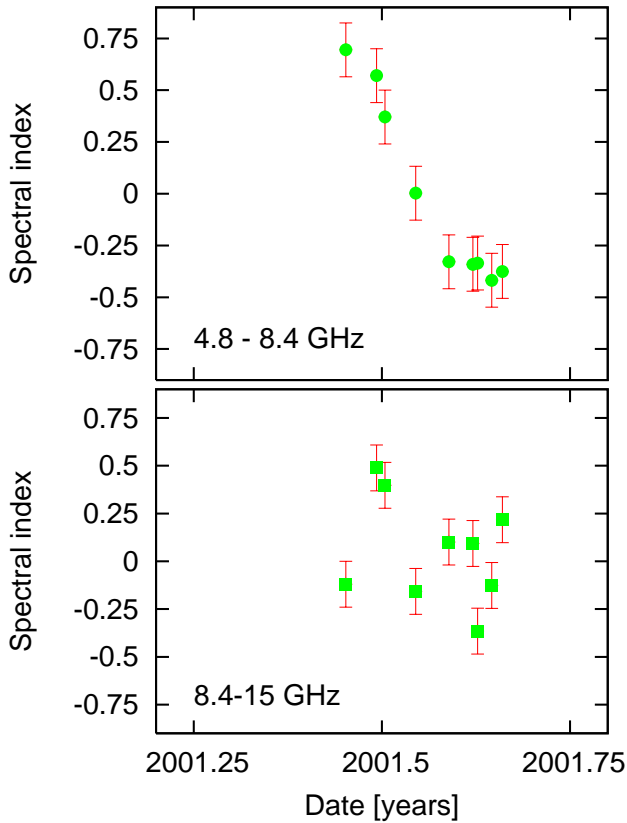}}
\caption{Spectral indices of M81* between 4.8 and 8.4 GHz (top) and 8.4 and 15 GHz (bottom) during the flare in the new VLA observations. }
\label{flare-spec}
\end{figure}

M81* underwent a flare in total intensity during the new VLA observations 
between June and August 2001. The peak was reached at 8.4 GHz before 15 June 
2001, while the flux density continued to rise at 4.8 until 4 July 2001. At 15 
GHz, the peak was reached in 30 June 2001. Fig.~\ref{flare-spec} shows the 
spectral indices between 4.8 and 8.4 GHz, and bewteen 8.4 and 15 GHz during 
this flare. The spectrum at the lower frequencies shows a smooth transition 
from an inverted ($\alpha\sim$ +0.7) to a steep ($\alpha\sim$~-~0.4) spectrum.
At higher frequencies, the spectral index does not follow a trend and is 
scattered between -0.4 and +0.5. The fast change in spectral index between 
4.8 and 8.4 GHz could be caused by a drop in the turnover frequency of a 
syncrotron self-absorpted jet from above 8.4 to below 4.8 GHz and
should be accompanied by a fast expansion of the jet. This 
behaviour is known in other active galactic nuclei (e.g. III~Zw~2: 
\citeNP{BrunthalerFalckeBower2000}, \citeNP{BrunthalerFalckeBower2005}). 
The scatter of the spectral index between 8.4 and 15 GHz could be caused by
multiple sub-flares that occur at 15 GHz.

\section{Discussion}

The origin of circular polarization in AGN is still not known. 
Several mechanisms have been proposed in the literature. 
Interstellar propagation effects predict a very steep spectrum 
(\citeNP{MacquartMelrose2000}) which is not consistent with our observations.
 One possible mechanism could be Faraday conversion 
(\citeNP{Pacholczyk1977}; \citeNP{JonesODell1977}) of linear polarization to 
circular polarization caused by the lowest energy relativistic electrons. 
\citeN{BowerFalckeBacker1999} proposed a simple model for Sgr~A* in which 
low-energy electrons reduce linear polarization through Faraday de-polarization
and convert linear polarization into circular polarization. Faraday conversion
can also affect the spectral properties of circular polarization and may
lead to a variety of spectral indices, including inverted spectra
(\citeNP{JonesODell1977}). In inhomogeneous sources, conversion can
produce relatively high fractional circular polarization (\citeNP{Jones1988}).
Gyro-synchrotron emission, can also lead to high circular polarization
with an inverted spectrum and low linear polarization (\citeNP{Ramaty1969}). 
However, this mechanism is to some degree related and also requires that
M81* and Sgr A* both contain a rather large number of low-energy
electrons. Faraday conversion is also favored by \citeN{BeckertFalcke2002} and
\citeN{RuszkowskiBegelman2002}.

%The absence of linear polarization up to a frequency of 22 GHz implies a 
%minimum rotation measure for depolarization of $10^4 \rdm$ 
%(e.g. \citeNP{BowerBackerZhao1999}).

The long-term stability of the sign of circular polarization suggests that the
Faraday conversion is connected to fundamental properties of the source and the
material which is responsible for the conversion. It requires uniformity in the
magnetic pole and accretion conditions over the observed timescales. One 
possible scenario is described by \citeN{ensslin2003} where the sign of the 
circular polarization is connected to the sense of rotation of the central 
engine. In this scenario M81* is expected to rotate counter-clockwise. 

The size of the radio emission of M81* at 8.4 GHz is $\sim 0.45$ mas, or 
$\sim$ 1800 AU at 4 Mpc (\citeNP{BietenholzBartelRupen1996}). The fact that
M81* is depolarized at at level of $<$ 0.1$\%$ at the same frequency requires 
that the depolarizing material is also present at a scale of $\sim$ 1800 AU.

While the polarization properties of M81* and Sgr A* are strikingly similar 
the total intensity spectrum seems to be different. In Sgr A*, the radio flux 
density rises towards the sub-mm regime (e.g. 
\citeNP{ZylkaMezgerWard-Thompson1995}, \citeNP{FalckeMarkoff2000}). Our
measurements indicate a different trend for M81*. Although the spectrum in 
Fig.~\ref{bb-spec} is not simultaneous and M81* shows strong variability 
our measured flux densities at 86 and 230 GHz are lower than the typical flux 
densities at centimeter wavelengths. This is different from the results in 
\citeN{ReuterLesch1996} who find an inverted spectrum up to $\sim$ 100 GHz.
We can not tell whether we observed M81* in a phase with unusual low millimeter 
emission or the \citeN{ReuterLesch1996} observations were made during an 
outburst at millimeter wavelengths. A simultaneous monitoring project from 
centimeter to millimeter wavelengths would be needed to decide this question.

Linear polarization is not detected for M81* at any wavelength
longward of 3.6 mm.  The presence of a jet at VLBI resolution,
radio synchrotron emission, and circular polarization suggest
that M81* is intrinsically linearly polarized but depolarized
during propagation through a magnetized plasma.  The case
is similar to that of Sgr A*, which is detected in linear polarization
only at wavelengths shortward of 3.6 mm.  For Sgr A*, the detection
of linear polarization at short wavelengths provides an upper limit
to the rotation measure of a few times $10^6 {\rm\ rad\ m^{-2}}$.
This provides an upper limit to the accretion rate of $\sim 10^{-7}
M_\odot y^{-1}$.  For M81*, we find a lower limit to the rotation
measure of $\sim 10^4 \rdm$ for the case of beam depolarization
and $\sim 4\times 10^5 \rdm$ for the case of bandwidth depolarization,
under the assumption that the intrinsic source is polarized.  The
lesser value could originate in the dense interstellar medium but
the larger value exceeds that seen anywhere in the ISM.  ADAF
and Bondi-Hoyle accretion models will depolarize the source unless
the accretion rate falls below $10^{-9} M_\odot y^{-1}$
(\citeNP{QuataertGruzinov2000}).  However, for radiatively
inefficient accretion flows, the larger of the RM limits implies
a lower limit to the accretion rate of $10^{-7} M_\odot y^{-1}$.
The accretion rate necessary for the X-ray luminosity is $10^{-5}
M_\odot y^{-1}$.  Since bandwidth depolarization effects decrease
as $\lambda^3$, measurement of the linear polarization at
a wavelength of 0.8 mm would increase the accuracy of the accretion
rate constraint by nearly two orders of magnitude.

\section{Summary \& Conclusion}

We have presented VLA observations of M81* from 1994 until 2002
that show that circular polarization is present at 4.8, 8.4, and 15
GHz in absence of linear polarization. The fractional circular
polarization is variable on timescales of days and months and not
correlated with the total flux density of the source. The sign of the
circular polarization was, if detected, at all frequencies and times always 
positive. The polarization properties are strikingly similar to the 
properties of Sgr A*, the central radio source in the Milky Way. This 
supports the hypothesis that M81* is a scaled up version of Sgr A*.
\citeN{AitkenGreavesChrysostomou2000} and \citeN{BowerWrightFalcke2003} 
detected linear polarization at 230 GHz and higher frequencies that also shows
variability (\citeNP{BowerFalckeWright2005}; \citeNP{MarroneMoranZhao2006}). 
Given the similarity between M81* and Sgr A* we expect to see also linear 
polarization in M81* at higher frequencies.

\begin{acknowledgements}
This research was partially supported by the DFG Priority Programme 1177. The 
National Radio Astronomy Observatory is operated by Associated Universities, 
Inc., under a cooperative agreement with the National Science Foundation.
\end{acknowledgements}

\bibliography{brunthal_refs}
\bibliographystyle{aa}

\appendix{}

\section{Tables}

\begin{table*}
\begin{center}
\caption{Circularly polarized flux at 4.8 GHz for M81* and calibrators. The errors on the fractional circular polarization are separated into statistical and systematic terms for the target and check source. For the calibrator source only the statistical error is given.\label{all-c}}
\begin{tabular}{rrrrrr}
\hline\hline
Date & Source & I  & $P_{c}$ & rms & $m_{c}$\\
& & [mJy] &[mJy] &[mJy] & $[\%]$\\
\hline
28 Jan. 1994 & 0954+658 &    622.2  & $<$ 0.43 &0.11 & $<$ 0.07 $\pm$ 0.02\\
             & M81*     &     95.0  &     0.36 &0.06 &     0.38 $\pm$ 0.06
$\pm$ 0.04\\
             & SN1993J  &     77.1  & $<$ 0.17 &0.05 & $<$ 0.22 $\pm$ 0.06
$\pm$ 0.03\\
\hline
21 Apr. 1994 & 0954+658 &    534.7  & $<$ 0.21 &0.07 & $<$ 0.04 $\pm$ 0.01\\
             & M81*     &    114.1  &     0.52 &0.05 &     0.46 $\pm$ 0.04
$\pm$ 0.03\\
             & SN1993J  &     62.1  & $<$ 0.22 &0.04 & $<$ 0.35 $\pm$ 0.06
$\pm$ 0.03\\
\hline
29 Aug. 1994 & 0954+658 &    597.2  & $<$ 0.18 &0.06 & $<$ 0.03 $\pm$ 0.01\\
             & M81*     &    109.0  &     0.21 &0.04 &     0.19 $\pm$ 0.04
$\pm$ 0.03\\
             & SN1993J  &     54.6  &     0.18 &0.03 &     0.33 $\pm$ 0.05
$\pm$ 0.03\\
\hline
\hline
02 Mar. 2001 & 1044+719 & 1568.2 & 0.11 & 0.11 & $<$ 0.01 $\pm$ 0.01\\
             & M81* & 130.5 & 0.44 & 0.10 & 0.34  $\pm$ 0.08 $\pm$0.05\\
             & 1053+704 & 388.2 & 0.03 & 0.07 & $<$ 0.01  $\pm$ 0.02  $\pm$0.06\\
\hline
15 Jun. 2001 & 1044+719 & 1710.9 & 0.33 & 0.35 & $<$ 0.02  $\pm$ 0.02 \\
             & M81* & 136.7 & 0.45 & 0.11 & 0.33  $\pm$ 0.08 $\pm$0.05\\
             & 1053+704 & 490.1 & -0.58 & 0.21 & $<$ 0.12  $\pm$ 0.04 $\pm$0.06\\
\hline
30 Jun. 2001 & 1044+719 & 1645.4 & 0.4 & 0.32 & $<$ 0.02  $\pm$ 0.02\\
             & M81* & 140.9 & 0.3 & 0.12 & $<$ 0.21  $\pm$ 0.09 $\pm$0.05\\
             & 1053+704 & 497.5 & -1.41 & 0.33 & $<$ 0.28  $\pm$ 0.07 $\pm$0.07\\
\hline
04 Jul. 2001 & 1044+719 & 1687.5 & 0.01 & 0.01 & $<$ 0.01  $\pm$ 0.01\\
             & M81* & 142.8 & 0.12 & 0.05 & $<$ 0.09  $\pm$ 0.04 $\pm$0.04\\
             & 1053+704 & 484 & 0.01 & 0.01 & $<$ 0.01  $\pm$ 0.01  $\pm$0.05\\
\hline
19 Jul. 2001 & 1044+719 & 1645.3 & 0.08 & 0.24 & $<$ 0.01  $\pm$ 0.05\\
             & M81* & 134.9 & 0.46 & 0.19 & $<$ 0.34  $\pm$ 0.14 $\pm$0.07\\
             & 1053+704 & 516.8 & 0.12 & 0.12 & $<$ 0.02  $\pm$ 0.02 $\pm$0.10\\
\hline
04 Aug. 2001 & 1044+719 & 1641.8 & 0.1 & 0.31 & $<$ 0.01  $\pm$ 0.02\\
             & M81* & 128.6 & 0.07 & 0.07 & $<$ 0.05  $\pm$ 0.05 $\pm$0.07\\
             & 1053+704 & 564.3 & 0.14 & 0.14 & $<$ 0.02  $\pm$ 0.02 $\pm$0.10\\
\hline
16 Aug. 2001 & 1044+719 & 1636.5 & 0.06 & 0.23 & $<$ 0.01  $\pm$0.01 \\
             & M81* & 122.7 & 0.44 & 0.18 & $<$ 0.36  $\pm$ 0.15 $\pm$0.07\\
             & 1053+704 & 524.3 & -0.38 & 0.23 & $<$ 0.07  $\pm$0.04  $\pm$0.10 \\
\hline
18 Aug. 2001 & 1044+719 & 1607.5 & 0.01 & 0.01 & $<$ 0.01  $\pm$ 0.01\\
             & M81* & 118.2 & 0.74 & 0.16 & 0.63  $\pm$ 0.14  $\pm$ 0.07\\
             & 1053+704 & 532.6 & 0.14 & 0.14 & $<$ 0.03  $\pm$ 0.03 $\pm$ 0.10\\
\hline
25 Aug. 2001 & 1044+719 & 1602.7 & 0.02 & 0.15 & $<$ 0.01  $\pm$ 0.01\\
             & M81* & 112.7 & 0.37 & 0.16 & $<$ 0.33  $\pm$ 0.14 $\pm$ 0.07\\
             & 1053+704 & 546 & 0.13 & 0.13 & $<$ 0.02  $\pm$ 0.02 $\pm$ 0.10\\
\hline
30 Aug. 2001 & 1044+719 & 1655.7 & 0.13 & 0.13 & $<$ 0.01  $\pm$ 0.01\\
             & M81* & 113.9 & 0.23 & 0.11 & $<$ 0.2  $\pm$  0.10 $\pm$ 0.07\\
             & 1053+704 & 548 & -0.08 & 0.14 & $<$ 0.01  $\pm$ 0.03 $\pm$ 0.10\\
\hline

\end{tabular}
\end{center}
\end{table*}

\begin{table*}
\begin{center}
\caption{Circularly polarized flux at 8.4 GHz for M81* and calibrators. The errors on the fractional circular polarization are separated into statistical and systematic terms for the target and check source. For the calibrator source only the statistical error is given.\label{all-x}}
\begin{tabular}{rrrrrr}
\hline\hline
Date & Source & I  & $P_{c}$ & rms & $m_{c}$\\
& & [mJy] &[mJy] &[mJy] & $[\%]$\\
\hline
05 Nov. 1993 & 0954+658 &   663.5  & $<$ 0.27 &0.04&  $<$ 0.04 $\pm$ 0.01\\
             & M81*     &   110.4  &     0.59 &0.03&      0.53 $\pm$ 0.03
$\pm$ 0.04\\
             & SN1993J  &    62.2  & $<$ 0.07 &0.02&  $<$ 0.11 $\pm$ 0.03
$\pm$ 0.03\\
\hline
16 Dec. 1993 & 0954+658 &   655.5  & $<$ 0.12 &0.04&  $<$ 0.02 $\pm$ 0.01\\
             & M81*     &    85.7  &     0.23 &0.03&      0.27 $\pm$ 0.04 
$\pm$ 0.03\\
             & SN1993J  &    54.7  & $<$ 0.06 &0.02&  $<$ 0.11 $\pm$ 0.04
$\pm$ 0.03\\
\hline
28 Jan. 1994 & 0954+658 &   600.6  & $<$ 0.24 &0.08&  $<$ 0.04 $\pm$ 0.01 \\
             & M81*     &   111.2  &     0.79 &0.04&      0.71 $\pm$ 0.04
$\pm$ 0.04  \\
             & SN1993J  &    48.9  & $<$ 0.14 &0.03&  $<$ 0.29 $\pm$ 0.06 
$\pm$ 0.03\\
\hline
29 Aug. 1994 & 0954+658 &   648.1  & $<$ 0.11 &0.04&  $<$ 0.02 $\pm$ 0.01\\
             & M81*     &   102.0  &     0.35 &0.03&      0.34 $\pm$ 0.03
$\pm$ 0.04\\
             & SN1993J  &    34.5  & $<$ 0.08 &0.03&  $<$ 0.23 $\pm$ 0.09
$\pm$ 0.03\\
\hline
31 Oct. 1994 & 0954+658 &   670.9  & $<$ 0.12 &0.04&  $<$ 0.02 $\pm$ 0.01\\
             & M81*     &   117.2  &     0.33 &0.03&      0.28 $\pm$ 0.03
$\pm$ 0.04\\
             & SN1993J  &    33.5  & $<$ 0.08 &0.03&  $<$ 0.24 $\pm$ 0.09
$\pm$ 0.03\\
\hline
23 Dec. 1994 & 0954+658 &   693.7  & $<$ 0.43 &0.14&  $<$ 0.06 $\pm$ 0.02\\
             & M81*     &    76.7  &     0.52 &0.09&      0.68 $\pm$ 0.12
$\pm$ 0.07\\
             & SN1993J  &    30.4  & $<$ 0.22 &0.07&  $<$ 0.72 $\pm$ 0.23
$\pm$ 0.05\\
\hline
07 Apr. 1996 & 0954+658 &   786.3  & $<$ 0.11 &0.04& $<$ 0.01 $\pm$ 0.01\\
             & M81*     &   165.8  &     1.15 &0.03&     0.69 $\pm$ 0.02
$\pm$ 0.04\\
             & SN1993J  &    20.5  & $<$ 0.15 &0.05& $<$ 0.73 $\pm$ 0.24
$\pm$ 0.04\\
\hline
\hline
02 Mar. 2001 & 1044+719 & 1478.7 & 0.13 & 0.13 & $<$ 0.01   $\pm$ 0.01\\
             & M81* & 143.8 & 0.62& 0.07 & 0.43   $\pm$ 0.05  $\pm$ 0.05\\
             & 1053+704 & 524.8 & 0.35 & 0.10 & $<$ 0.07   $\pm$ 0.02  $\pm$ 0.06\\
\hline
15 Jun. 2001 & 1044+719 & 2136.1 & 0.11 & 0.11 & $<$ 0.01   $\pm$ 0.01\\
             & M81* & 201.7 & 1.08 & 0.10 & 0.53 $\pm$ 0.05  $\pm$ 0.05\\
             & 1053+704 & 1006.1 & 0.17 & 0.17 & $<$ 0.02 $\pm$ 0.02  $\pm$ 0.06\\
\hline
30 Jun. 2001 & 1044+719 & 1488.6 & 0.01 & 0.01 & $<$ 0.01   $\pm$ 0.01\\
             & M81* & 193.9 & 0.75 & 0.11 & 0.39   $\pm$ 0.06  $\pm$ 0.05\\
             & 1053+704 & 763.0 & 0.34 & 0.19 & $<$ 0.04   $\pm$ 0.02  $\pm$ 0.07\\
\hline
04 Jul. 2001 & 1044+719 & 1525.8 & 0.57 & 0.57 & $<$ 0.04   $\pm$ 0.04\\
             & M81* & 175.7 & 0.93 & 0.09 & 0.53   $\pm$ 0.05  $\pm$ 0.05\\
             & 1053+704 & 753.3 & 0.13 & 0.13 & $<$ 0.02   $\pm$ 0.02  $\pm$ 0.06\\
\hline
19 Jul. 2001 & 1044+719 & 1474.8 & 0.10 & 0.10 & $<$ 0.01   $\pm$ 0.01\\
             & M81* & 135.1 & 0.60 & 0.09 & 0.45   $\pm$ 0.07  $\pm$ 0.05\\
             & 1053+704 & 769.2 & 0.27 & 0.13 & $<$ 0.04   $\pm$ 0.02  $\pm$ 0.07\\
\hline
04 Aug. 2001 & 1044+719 & 1485.4 & 0.12 & 0.12 & $<$ 0.01   $\pm$ 0.01\\
             & M81* & 107.0 & 0.84 & 0.17 & 0.78   $\pm$ 0.16  $\pm$ 0.05\\
             & 1053+704 & 785.1 & 0.57 & 0.26 &  $<$ 0.07   $\pm$ 0.03  $\pm$ 0.07\\
\hline
16 Aug. 2001 & 1044+719 & 1469.1 & 0.27 & 0.27 & $<$ 0.02   $\pm$ 0.02\\
             & M81* & 101.4 & 0.36 & 0.16 & $<$ 0.35   $\pm$ 0.16  $\pm$ 0.07\\
             & 1053+704 & 758.3 & 1.20 & 0.27 & $<$ 0.16   $\pm$ 0.04  $\pm$ 0.10\\
\hline
18 Aug. 2001 & 1044+719 & 1476.8 & 0.41 & 0.65 & $<$ 0.03   $\pm$ 0.03\\
             & M81* & 98.0 & 0.40 & 0.09 & 0.41   $\pm$ 0.09  $\pm$ 0.07\\
             & 1053+704 & 775.2 & 0.41 & 0.30 & $<$ 0.05   $\pm$ 0.04   $\pm$ 0.10\\
\hline
25 Aug. 2001 & 1044+719 & 1406.9 & 0.11 & 0.11 & $<$ 0.01   $\pm$ 0.01\\
             & M81* & 89.2 & 0.32 & 0.09 & 0.36   $\pm$ 0.10  $\pm$ 0.05\\
             & 1053+704 & 746.0 & 1.30 & 0.24 & $<$ 0.17   $\pm$ 0.03  $\pm$ 0.07\\
\hline
30 Aug. 2001 & 1044+719 & 1520 & 0.17 & 0.17 & $<$ 0.01   $\pm$ 0.01\\
             & M81* & 92.3 & 0.42 & 0.16 & 0.45   $\pm$ 0.17 $\pm$  0.07\\
             & 1053+704 & 789.2 & 0.30 & 0.30 & $<$ 0.04   $\pm$ 0.04  $\pm$ 0.10\\
\hline
\end{tabular}
\end{center}
\end{table*}

\begin{table*}
\begin{center}
\caption{Circularly polarized flux at 15 GHz for M81* and calibrators. The errors on the fractional circular polarization are separated into statistical and systematic terms for the target and check source. For the calibrator source only the statistical error is given.\label{all-u}}
\begin{tabular}{rrrrrr}
\hline\hline
Date & Source & I  & $P_{c}$ & rms & $m_{c}$\\
& & [mJy] &[mJy] &[mJy] & $[\%]$\\
\hline
05 Nov. 1993 & 0954+658 &    664.5  & $<$ 0.26 &0.09& $<$ 0.04 $\pm$ 0.01\\
             & M81*     &    108.2  &     1.14 &0.18&     1.05 $\pm$ 0.17
$\pm$ 0.05\\
             & SN1993J  &     42.3  & $<$ 4.05 &1.35& $<$ 9.57 $\pm$ 3.19
$\pm$ 0.03\\
\hline
16 Dec. 1993 & 0954+658 &    606.4  & $<$ 0.28 &0.09& $<$ 0.05 $\pm$ 0.01\\
             & M81*     &     87.2  & $<$ 0.06 &0.19& $<$ 0.07 $\pm$ 0.22
$\pm$ 0.05\\
             & SN1993J  &     39.0  & $<$ 9.0  &3.32& $<$ 23.0 $\pm$ 8.51
$\pm$ 0.03\\
\hline
\hline
02 Mar. 2001 & 1044+719 & 999.8 & 0.03 & 0.12 & $<$ 0.01 $\pm$ 0.01\\
         & M81* & 114.2 & 0.81 & 0.19 & 0.71  $\pm$ 0.17 $\pm$ 0.05\\
         & 1053+704 & 485.2 & 0.25 & 0.18 & $<$ 0.05  $\pm$ 0.04 $\pm$ 0.07\\
\hline
15 Jun. 2001 & 1044+719 & 2275 & 0.36 & 0.36 & $<$ 0.02 $\pm$ 0.02\\
             & M81* & 188.1 & 2.16 & 0.38 & 1.15  $\pm$ 0.20 $\pm$ 0.06\\
             & 1053+704 & 1354.7 & 8.28 & 0.75 & 0.61  $\pm$ 0.06 $\pm$ 0.07\\
\hline
30 Jun. 2001 & 1044+719 & 1669.9 & 0.43 & 0.43 & $<$ 0.03  $\pm$ 0.03\\
             & M81* & 257.4 & 3.14 & 0.4 & 1.22  $\pm$ 0.16 $\pm$ 0.07\\
             & 1053+704 & 1051.5 & 2.61 & 0.66 & $<$ 0.25 $\pm$ 0.06 $\pm$ 0.09\\
\hline
04 Jul. 2001 & 1044+719 & 1955.8 & 1.67 & 1.67 & $<$ 0.09  $\pm$ 0.09 \\
             & M81* & 221.2 & 3.66 & 0.46 & 1.66  $\pm$ 0.21 $\pm$ 0.06\\
             & 1053+704 & 1162.1 & 4.9 & 0.88 & 0.42 $\pm$ 0.08  $\pm$ 0.07\\
\hline
19 Jul. 2001 & 1044+719 & 1663.3 & 0.58 & 0.58 & $<$ 0.03 $\pm$ 0.03\\
             & M81* & 123.3 & 1.65 & 0.68 & $<$ 1.34  $\pm$ 0.55 $\pm$ 0.09\\
             & 1053+704 & 1011.9 & 1.76 & 1.05 & $<$ 0.17  $\pm$ 0.10 $\pm$ 0.12\\
\hline
04 Aug. 2001 & 1044+719 & 1663 & 0.29 & 1.14 & $<$ 0.02  $\pm$ 0.07\\
             & M81* & 113.4 & 0.36 & 0.36 & $<$ 0.31  $\pm$ 0.31 $\pm$ 0.09\\
             & 1053+704 & 973.4 & 6.42 & 1.92 & $<$ 0.66  $\pm$ 0.20 $\pm$ 0.12\\
\hline
16 Aug. 2001 & 1044+719 & 1876 & 0.51 & 0.51 & $<$ 0.03  $\pm$ 0.03\\
             & M81* & 107 & 0.38 & 0.42 & $<$ 0.35  $\pm$ 0.39  $\pm$ 0.09\\
             & 1053+704 & 1043.9 & 3.42 & 1.37 & $<$ 0.33  $\pm$ 0.13  $\pm$ 0.12\\
\hline
18 Aug. 2001 & 1044+719 & 1496.6 & 0.46 & 0.98 & $<$ 0.03  $\pm$ 0.07\\
             & M81* & 79.3 & 0.2 & 0.2 & $<$ 0.25  $\pm$ 0.25 $\pm$ 0.09\\
             & 1053+704 & 830.4 & 2.44 & 0.95 & $<$ 0.29  $\pm$ 0.11  $\pm$ 0.12\\
\hline
25 Aug. 2001 & 1044+719 & 1378 & 0.49 & 0.49 & $<$ 0.04  $\pm$ 0.04\\
             & M81* & 82.9 & 0.47 & 0.76 & $<$ 0.56  $\pm$ 0.56 $\pm$ 0.09\\
             & 1053+704 & 756.7 & 0.82 & 1.65 & $<$ 0.11  $\pm$ 0.22 $\pm$ 0.12\\
\hline
30 Aug. 2001 & 1044+719 & 1846.3 & 0.76 & 0.76 & $<$ 0.04  $\pm$ 0.04\\
             & M81* & 104.7 & 0.54 & 0.53 & $<$ 0.52  $\pm$ 0.52 $\pm$ 0.09\\
             & 1053+704 & 988.9 & 0.41 & 0.41 & $<$ 0.04  $\pm$ 0.04 $\pm$ 0.12\\
\hline
\end{tabular}
\end{center}
\end{table*}

%\begin{table*}
%\begin{center}
%\caption{VLA observations of M81* on 09 August 2003. \label{VLAhigh}}
%\begin{tabular}{rccrrrrr}
%\hline\hline
%Date &   Frequency  & Sideband &   I  &     Q   &    U   &    V    &     m$_p$\\
%&[GHz]&&[mJy] & [mJy] &[mJy] &[mJy] &$[\%]$\\
%\hline
%09 Aug. 2003 &  15 & & 125.9 $\pm$ 1.5 & $<$ 0.45 & $<$ 0.45 & $<$ 0.45 &\\
%09 Aug. 2003 &  22 & & 118.2 $\pm$ 1.3 & $<$ 0.36 & $<$ 0.36 & $<$ 0.42 &\\
%09 Aug. 2003 &  43 & &  66.8 $\pm$ 2.0 & $<$ 1.1  & $<$ 1.1  & $<$ 1.4  &\\
%\hline
%\end{tabular}
%\end{center}
%\end{table*}

\begin{table*}
\begin{center}
\caption{Polarized and total flux density of M81* at high frequencies unsing the VLA (15, 22, and 43 GHz) and BIMA (83, 86, and 230 GHz).~\label{High}}
\begin{tabular}{rccrrrrr}
\hline\hline
Date & Frequency &  Sideband  &    I  &     Q   &    U   &    V    &     m$_p$\\
&[GHz]&&[mJy] & [mJy] &[mJy] &[mJy] &$[\%]$\\
\hline
09 Aug. 2003 &  15 & & 125.9 $\pm$ 1.5 & $<$ 1.4 & $<$ 1.4 & $<$ 1.4 & $<$ 1.0 \\
             &  22 & & 118.2 $\pm$ 1.3 & $<$ 1.1 & $<$ 1.1 & $<$ 1.3 & $<$ 1.0 \\
             &  43 & &  66.8 $\pm$ 2.0 & $<$ 3.3 & $<$ 3.3 & $<$ 4.2 & $<$ 4.9 \\
\hline
\hline
07 Sep. 2003 & 83 & lsb  & 44.0 $\pm$ 2.6 & -2.0 $\pm$ 2.6 &  5.6 $\pm$ 2.6 & 0.3 $\pm$ 2.6 &  \\
             & 86 & usb  & 41.8 $\pm$ 2.6 & -1.8 $\pm$ 2.6 & -3.4 $\pm$ 2.6 & 3.2 $\pm$ 2.6 &   \\
             & & avg   & 42.9 $\pm$ 1.8 & -1.8 $\pm$ 1.8 &  1.1 $\pm$ 1.8 & 1.8 $\pm$ 1.8 &  5.1 $\pm$ 4.2 \\
\hline
12 Sep. 2003 & 83 & lsb  & 89.4 $\pm$ 1.8 & -1.3 $\pm$ 1.8 & -2.1 $\pm$ 1.8 & 5.7 $\pm$ 1.8 &   \\
             & 86 & usb  & 92.4 $\pm$ 1.8 & -0.8 $\pm$ 1.8 & -1.1 $\pm$ 1.8 & 2.5 $\pm$ 1.8 &   \\
             & & avg   & 90.9 $\pm$ 1.3 & -1.1 $\pm$ 1.3 & -1.6 $\pm$ 1.3 & 4.1 $\pm$ 1.3 &  2.1 $\pm$ 1.4 \\
\hline
21 Sep. 2003 & 83 & lsb  & 86.4 $\pm$ 1.5 & -0.2 $\pm$ 1.5 & -0.7 $\pm$ 1.5 & 3.8 $\pm$ 1.5 &   \\
             & 86 & usb  & 86.9 $\pm$ 1.5 & -1.3 $\pm$ 1.5 & -0.2 $\pm$ 1.5 & 3.4 $\pm$ 1.5 &   \\
             & & avg   & 86.7 $\pm$ 1.1 & -0.2 $\pm$ 1.1 & -0.5 $\pm$ 1.1 & 3.6 $\pm$ 1.1 &  0.6 $\pm$ 1.3 \\
\hline
06 Oct. 2003 & 83 & lsb  & 70.4 $\pm$ 2.0 &  3.5 $\pm$ 2.0 &  0.7 $\pm$ 2.0 & 0.2 $\pm$ 2.0 &   \\
             & 86 & usb  & 72.7 $\pm$ 2.0 & -0.3 $\pm$ 2.0 &  0.3 $\pm$ 2.0 & 3.5 $\pm$ 2.0 &   \\
             & & avg   & 71.6 $\pm$ 1.4 &  1.6 $\pm$ 1.4 &  0.5 $\pm$ 1.4 & 1.8 $\pm$ 1.4 &  2.3 $\pm$ 2.0 \\
\hline
09 Oct. 2003 & 83 & lsb  & 46.1 $\pm$ 1.8 &  5.1 $\pm$ 1.8 & -2.1 $\pm$ 1.8 & 1.0 $\pm$ 1.8 &  \\
             & 86 & usb  & 45.1 $\pm$ 1.8 & -3.2 $\pm$ 1.8 &  2.7 $\pm$ 1.8 & 1.5 $\pm$ 1.8 &   \\
             & & avg   & 45.6 $\pm$ 1.3 &  1.0 $\pm$ 1.3 &  0.3 $\pm$ 1.3 & 1.2 $\pm$ 1.3 &  2.2 $\pm$ 2.9 \\
\hline
12 Oct. 2003 & 83 & lsb  & 43.9 $\pm$ 1.5 & -2.4 $\pm$ 1.5 & -1.0 $\pm$ 1.5 & 3.8 $\pm$ 1.5 &   \\
             & 86 & usb  & 34.5 $\pm$ 1.5 &  6.1 $\pm$ 1.5 &  4.9 $\pm$ 1.5 &-1.0 $\pm$ 1.5 &  \\
             & & avg   & 39.2 $\pm$ 1.1 &  1.9 $\pm$ 1.1 &  2.0 $\pm$ 1.1 & 1.4 $\pm$ 1.1 &  7.0 $\pm$ 2.8 \\
\hline
\hline
01 Nov. 2003 & 230 & lsb  &  33.7 $\pm$  6 & -6.6 $\pm$  6 &   1.2 $\pm$  6 &  -4.5 $\pm$  6 & \\
             &     & usb  &  27.5 $\pm$  6 &  2.7 $\pm$  6 & -19.9 $\pm$  6 &   7.3 $\pm$  6 & \\
             &     & avg  &  30.6 $\pm$  4 & -2.0 $\pm$  4 &  -9.4 $\pm$  4 &   1.4 $\pm$  4 & 31.4 $\pm$ 13\\
\hline
\end{tabular}
\end{center}
\end{table*}

\end{document}